\documentclass[11pt]{article}
\usepackage{cite}
\usepackage{amsmath,amsfonts,amssymb}
\usepackage[small,bf,hang]{caption}

\def\hybrid{
        \topmargin -20pt
        \oddsidemargin 0pt
        \headheight 0pt \headsep 0pt
        \textwidth 6.25in 
        \textheight 9.5in 
        \marginparwidth .875in
        \parskip 5pt plus 1pt \jot = 1.5ex}

\hybrid

 \csname
@addtoreset\endcsname{equation}{section}

\newcommand{\vev}[1]{\langle #1 \rangle}

\def\moth{\mathsurround=0pt}
\newdimen\zo \zo=0pt

\def\tick{\leaders\hrule height 0.5ex depth 0pt \hskip 0.5pt}
\def\upboxfill{$\moth \setbox\zo\hbox{\tick}%
  \hskip 3pt\hbox to 0pt{$\tick$\hss}\hrulefill \hbox to 7.5pt{$\tick$\hss}$}

\def\dtick{\leaders\hrule height .34pt depth 0.5ex \hskip 0.5pt}
\def\downboxfill{$\moth \setbox\zo\hbox{\dtick}%
  \hskip 2pt\hbox to 0pt{$\dtick$\hss}\hrulefill \hbox to 2pt{$\dtick$\hss}$}

\def\bec{\begin{center}}
\def\ec{\end{center}}

\def\be{\begin{equation}}
\def\ee{\end{equation}}
\def\bea{\begin{eqnarray}}
\def\eea{\end{eqnarray}}
\def\ba{\begin{array}}
\def\ea{\end{array}}

\begin{document}

\begin{titlepage}
\rightline{February 2011}
\rightline{\tt MIT-CTP-4218}
\begin{center}
\vskip 2.5cm
{\Large \bf {
On factorizations in perturbative quantum gravity}}\\
\vskip 2.0cm
{\large {Olaf Hohm\hskip-4pt}}
\vskip 0.5cm
{\it {Center for Theoretical Physics}}\\
{\it {Massachusetts Institute of Technology}}\\
{\it {Cambridge, MA 02139, USA}}\\
ohohm@mit.edu

\vskip 2.5cm
{\bf Abstract}

\end{center}

\vskip 0.2cm
\noindent
\begin{narrower}

Some features of Einstein gravity are most easily understood from string theory but 
are not manifest at the level of the usual Lagrangian formulation. 
One example is the factorization of gravity amplitudes into gauge theory amplitudes.
Based on the recently constructed `double field theory' and
a geometrical frame-like formalism developed by Siegel, we provide a framework
of perturbative Einstein gravity coupled to a 2-form and a dilaton in which, as a consequence of
T-duality, the Feynman rules factorize to all orders in perturbation theory.
We thereby establish the precise relation between the field variables in different 
formulations and discuss the Lagrangian that, when written in terms of these variables, makes 
a left-right factorization manifest.

\end{narrower}

\end{titlepage}

\section{Introduction}\setcounter{equation}{0}
The usual perturbative approach of covariant quantum gravity on flat space starts
with the Einstein-Hilbert theory and
expands the Riemannian metric $g_{ij}$ around a constant background,
$g_{ij}=\delta_{ij}+h_{ij}$.
The diffeomorphism invariance then translates into a gauge symmetry of the
fluctuation $h_{ij}$, which to lowest order reads
$\delta h_{ij}=\partial_{i}\xi_{j}+\partial_{j}\xi_{i}$. From this point of view, the problem of formulating
the corresponding quantum field theory is conceptually not much different
from Yang-Mills theory.  Technically, however, gravity appears to be much more complicated than
Yang-Mills theory in that the Einstein-Hilbert action is non-polynomial in
the fluctuation $h_{ij}$ (and, of course, physically it is also very different in that it is
non-renormalizable).

In recent years it has become clear, however, that the amplitudes reveal a hidden
simplicity that is obscured at the level of the Lagrangian and the corresponding Feynman
rules; see \cite{Bern:2002kj,Bern:2008qj,ArkaniHamed:2008gz} and references therein.
For instance, string theory exhibits the so-called KLT relations
which imply a factorization of closed string or graviton amplitudes into open string or gauge
theory amplitudes \cite{Kawai:1985xq}. It turns out that similar KLT relations also hold in the
corresponding (low-energy) field theories, i.e., in Einstein gravity and its supersymmetric extensions. 
These relations have been instrumental in the
recent UV-finiteness proofs for ${\cal N}=8$ supergravity at higher loops; see
\cite{Dixon:2010gz} for a review.

Given these intriguing simplifications, it is natural to ask whether there is a way to make
these  properties, at least to some extent, manifest at the level of the Lagrangian.
Specifically, one has the freedom to perform field redefinitions of $h_{ij}$, and
one may expect that there is a field-basis that is better adapted to the
features inherited from string theory.
In fact, in closed string field theory, for instance, a non-linear and non-polynomial field
redefinition is required in order to connect the `string variables' to the `Einstein variables' $h_{ij}$
\cite{Michishita:2006dr}.

Early attempts to render the KLT relations manifest upon field redefinitions (and
non-linear generalizations of the de Donder gauge-fixing condition) are due to Bern and Grant
\cite{Bern:1999ji}; see \cite{ArkaniHamed:2008yf,Bern:2010yg} for more recent results.
The idea is to factorize the metric fluctuation into two `gauge vectors',
 \be
  h_{ij}\,\rightarrow\, A_{i}\,\bar{A}_{j}\;,
 \ee
and then to require that the Feynman rules factorize into `left-handed' parts depending only on
$A$ and `right-handed' parts depending only on $\bar A$. Put differently, one can think of the
first index of $h_{ij}$ as left-handed or unbarred and the second index as right-handed
or barred, and then demand that the Lagrangian contains only like-wise index contractions.
This requires a field redefinition which reads to lowest order \cite{Bern:1999ji}
 \be\label{fieldred}
  h_{ij} \,\rightarrow \, h_{ij}+\frac{1}{2}h_{i}{}^{k}h_{kj}+\cdots \;.
 \ee
Of course, since $h_{ij}$ is actually symmetric, this assignment of left and right indices
may sound somewhat obscure, in particular, this requirement is not very strong.
One purpose of this note is to introduce field redefinitions
that establish consistent left-right factorization to all orders in perturbation theory
for the low-energy theory of the bosonic string, i.e., for Einstein gravity coupled to a 2-form and a dilaton.
The relevant formulation allows to combine the metric and 2-form fluctuations into a
non-symmetric field $e_{ij}$, which is the natural variable in string theory and which leads to an
unambiguous  assignment of left-right indices.

To this end, we use the recently formulated `double field theory' and its T-duality invariance
\cite{Hull:2009mi,Hull:2009zb,Hohm:2010jy,Hohm:2010pp} (see also
\cite{Kwak:2010ew,Hohm:2010xe} and \cite{Hohm:2011gs} for a review).
Specifically, in this theory the space-time coordinates are doubled in such a way that the
T-duality group $O(D,D)$ (with $D$ denoting the space-time dimension) acts naturally,
and it is equivalent to the standard low-energy action when the dependence on the
new coordinates is dropped. The theory requires a constraint that eliminates
half of the coordinates, and
for the purposes of this paper we may thus think of the new coordinates as purely auxiliary
objects; in particular, we do not
require the coordinates to be compact.
We introduce this formulation in sec.~2 and review how the $O(D,D)$ invariance
of the full non-linear background independent action is, in fact,  equivalent to consistent
left-right index contractions (in a sense to be made precise below).
When expanded around a constant background, however, this left-right factorization
is no longer manifest in the sense required above, but the field redefinition that relates to the
basis in string field theory and that should therefore restore
this property is known to all orders \cite{Hull:2009zb,Michishita:2006dr}.
Here, we
prove that in this field basis the $O(D,D)$
invariance indeed implies left-right factorization to all orders in perturbation theory, thereby
solving the problem stated above.

Apart from the last technical step, this result is already largely contained
in the existing literature on double field theory,  but it
can actually be cast into a somewhat more geometrical language, using a powerful formalism
based on enlarged frame fields introduced
by Siegel \cite{Siegel:1993th,Siegel2}.
The frame field $e_{A}{}^{M}$ is subject to global $O(D,D)$ transformations, acting
in the fundamental representation indicated by indices $M,N,\ldots=1,\ldots,2D$,  and to local
$GL(D)\times GL(D)$ tangent space transformations, corresponding to the flat index
$A=(a,\bar a)$.\footnote{For an alternative formulation of gravity with local 
$GL(D)$ symmetry see \cite{Percacci:1990wy}.}
Siegel also introduces a perturbative expansion,
in which the fluctuation $h$ carries only flat indices,
 \be\label{ehexp}
  e_{A}{}^{M} \ = \ \langle e_{A}{}^{M}\rangle -h_{A}{}^{B}\langle e_{B}{}^{M}\rangle\;,
 \ee
where  $\langle e_{A}{}^{M}\rangle$ denotes the constant background. The tangent
space symmetry can be gauge-fixed such that the only independent variable is the
off-diagonal component $h_{a\bar b}$ \cite{Siegel:1993th}.

Recently, this frame formalism has been related in detail to the double field theory
and thereby to the conventional variables in Einstein gravity \cite{Hohm:2010pp,Hohm:2010xe}.
Based on this, we present in sec.~3 as the main results of this note
the precise relation between the perturbations in
Einstein gravity and string field theory on the one hand and
the perturbations in Siegel's frame formalism on the other.
Remarkably, we find that the string field theory variable, here denoted by $e_{ij}$,
can be identified with the frame-like variable $h_{a\bar b}$ \textit{to all orders},
 \be
  e_{ij} \ = \ \vev{e_{i}{}^{a}}\,\vev{e_{j}{}^{\bar b}}\,h_{a\bar b}\;.
 \ee
Here, the two (independent) background vielbeins can be rotated into Kronecker symbols
by means of background $GL(D)\times GL(D)$ transformations.
This formulation provides therefore a significant
technical simplification in that the field redefinition that establishes the left-right factorization
need not be carried out explicitly, but rather is implicitly incorporated by use of the frame-like variable.
It has already been pointed out by Siegel that this formalism allows us to make
certain features inherited from string theory manifest in conventional field theory \cite{Siegel:1993th}.
Here, this will be investigated explicitly, in particular we discuss the Lagrangian formulation 
that makes the left-right factorization manifest, as   
displayed in eq.~(\ref{newframeaction}) below.

\section{T-duality and redefinition of Einstein variables}
In this section we review the double field theory
and its $O(D,D)$ invariance. Next, using this duality invariance,
we prove that in the field-basis suggested by string field theory left-right
factorization is realized to all orders.

\subsection{Double field theory and Einstein variables}
We start from the standard low-energy action for bosonic string theory, i.e., with Einstein gravity
coupled to a 2-form $b_{ij}$ and a scalar dilaton $\phi$,
 \be\label{gravaxion}
  S \ = \ \int dx \sqrt{g}e^{-2\phi}\left[R+4(\partial\phi)^2-\frac{1}{12}H^2\right]\,,
 \ee
where $H_{ijk}=3\partial_{[i}b_{jk]}$.
The double field theory extension of this action is written in terms of a variable that combines the metric
and $b$-field into a `non-symmetric metric', ${\cal E}_{ij}=g_{ij}+b_{ij}$, and
a dilaton $d$, which is a density rather than a scalar and defined by $\sqrt{g} e^{-2\phi} = e^{-2d}$
\cite{Hohm:2010jy},
  \bea
  \label{THEAction}
  \begin{split}\hskip-10pt
  S \ = \ \int \,dx d\tilde{x}~
  e^{-2d}\Big[&
  -\frac{1}{4} \,g^{ik}g^{jl}   \,   {\cal D}^{p}{\cal E}_{kl}\,
  {\cal D}_{p}{\cal E}_{ij}
  +\frac{1}{4}g^{kl} \bigl( {\cal D}^{j}{\cal E}_{ik}
  {\cal D}^{i}{\cal E}_{jl}  + \bar{\cal D}^{j}{\cal E}_{ki}\,
  \bar{\cal D}^{i}{\cal E}_{lj} \bigr)~
  \\ &    + \bigl( {\cal D}^{i}\hskip-1.5pt d~\bar{\cal D}^{j}{\cal E}_{ij}
 +\bar{{\cal D}}^{i}\hskip-1.5pt d~{\cal D}^{j}{\cal E}_{ji}\bigr)
 +4{\cal D}^{i}\hskip-1.5pt d \,{\cal D}_{i}d ~\Big]\;,
 \end{split}
 \eea
where all indices are raised with $g^{ij}$, which is the inverse of $g_{ij}={\cal E}_{(ij)}$.
Here, all fields depend on the `doubled'  coordinates $X\equiv (x,\tilde{x})$, with derivatives defined
by
 \be\label{callder}
  {\cal D}_{i} \ = \ \frac{\partial}{\partial x^i}-{\cal E}_{ik}\frac{\partial}{\partial\tilde{x}_{k}}\;, \qquad
  \bar{\cal D}_{i} \ = \ \frac{\partial}{\partial x^i}+{\cal E}_{ki}\frac{\partial}{\partial\tilde{x}_{k}}\;.
 \ee
The consistency of the action requires the constraint
 \be\label{strongconstr}
  {\cal D}^{i}A\,{\cal D}_{i}B- \bar{\cal D}^{i}A\,\bar{\cal D}_{i}B \ = \ 0
 \ee
for arbitrary fields and parameters $A,B$, which implies that locally the fields depend only on
half of the coordiantes.
When the fields are assumed to be independent of $\tilde{x}$, i.e., $\tilde{\partial}=0$,
(\ref{THEAction}) is equivalent to (\ref{gravaxion}) \cite{Hohm:2010jy}.

The crucial property of the double field theory action (\ref{THEAction}) that will be used below is
its global $O(D,D)$ invariance under
 \begin{equation}\label{ETdual}
  {\cal E}^{\prime}(X^{\prime}) \ = \ (a{\cal E}(X)+b)(c{\cal E}(X)+d)^{-1}\;,
  \quad
  d^{\prime}(X^{\prime}) \ = \ d(X)\;,
 \end{equation}
where the coordinates transform in the fundamental representation of $O(D,D)$,
 \be
     X^{\prime} \ = \ \begin{pmatrix} \tilde{x}^{\prime} \\ x^{\prime} \end{pmatrix}
     \ = \  \begin{pmatrix} a &   b \\ c & d \end{pmatrix}
      \begin{pmatrix} \tilde{x} \\ x \end{pmatrix} \;,
      \qquad
  \begin{pmatrix} a &   b \\ c & d \end{pmatrix} \ \in \ O(D,D)\;.
 \ee
This invariance is not manifest, but can be reduced to consistent index contractions as
follows \cite{Hohm:2010jy}. The transformation behavior of the metric $g$ and the
calligraphic derivatives of ${\cal E}$ and $d$ is governed by the matrices ${\cal M}$  and
$ \bar{\cal M}$ defined by
 \be\label{MbarM}
   {\cal M}(X) \ = \ d^t-{\cal E}(X)c^t \;, \qquad \bar{\cal M}(X) \ = \ d^t+{\cal E}^t(X)c^t\;.
 \ee
More precisely, the (inverse) metric transforms as
\be
\label{fsetmatvmvgn9998}
 g^{ i j} \ = \  (\bar {\cal M}^{-1})_{p}{}^{i}\,
 g^{\prime\,  p q}  \,(\bar {\cal M}^{-1})_{q}{}^{ j}\,,~~~
 g^{ij} \ = \ ( {\cal M}^{-1})_{p}{}^{i} \, g^{\prime \,pq}\, ({\cal M}^{-1})_q{}^{j} \,,
\ee
and the calligraphic derivatives as
 \be
\label{vmswp}
{\cal D}_i  \ = \  {\cal M}_{i} {}^k \,{\cal D}'_k \,,  ~~~~
\bar{\cal D}_i  \ = \  \bar {\cal M}_{i}{}^k \, \bar{\cal D}'_k \,.
\ee
Moreover, despite the non-linear form of (\ref{ETdual}), it can be checked
that ${\cal E}$, acted on by a calligraphic derivative, transforms as
 \be
 {\cal D}_i{\cal E}_{jk} \ = \ {\cal M}_{i}^{~q}\,{\cal M}_{j}^{~p}\, \bar {\cal M}_{k}^{~\ell} \,\,
 {\cal D}'_q {\cal E}'_{p\ell} \,,
 ~~~\bar{\cal D}_i{\cal E}_{jk} \ = \ \bar {\cal M}_{i}^{~q}{\cal M}_{j}^{~p}\,
 \bar {\cal M}_{k}{}^{\ell}\,\bar{\cal D}_q^{\prime} {\cal E}'_{p\ell} \,.
\ee
Therefore, in this formalism there are two types of indices, unbarred and barred, corresponding to
a transformation with ${\cal M}$ or $\bar{\cal M}$ under $O(D,D)$. In particular, ${\cal E}_{ij}$ acted on by
a calligraphic derivative can be viewed as an object for which the first index is unbarred and the second
index is barred, while the index of an (un-)barred calligraphic derivative is (un-)barred.
Finally, from (\ref{fsetmatvmvgn9998}) we infer that the indices of the (inverse) metric can be
thought of as either both unbarred or both barred. The $O(D,D)$ invariance of (\ref{THEAction})
then follows from the fact that the action has only like-wise index contractions,
which one may easily confirm by inspection.

Thus, the double field theory formulation of the low-energy action exhibits already a left-right
factorization that it reminiscent to the requirement stated in the introduction. However,
once we consider the fluctuations around a flat background,
this factorization does not translate into a corresponding factorization in terms of the fluctuation.
To see this, we decompose ${\cal E}$ according to
 \be
  {\cal E}_{ij} \ = \ E_{ij}+\check{e}_{ij}\;,
 \ee
where we denoted the fluctuation by $\check{e}_{ij}$ (which is the sum of the usual metric
fluctuation $h_{ij}$ and the fluctuation of the 2-form) in order to distinguish it from the string field
theory variable $e_{ij}$ to be discussed below. Moreover, $E_{ij}=G_{ij}+B_{ij}$ encodes
the constant background metric and B-field. If one computes, for instance,
the inverse metric $g^{ij}$, it will contain arbitrary higher powers of $\check{e}_{(ij)}$ that generally
mix left- and right-indices.  Next, we discuss the field redefinition to $e_{ij}$ and show
that it restores the required left-right factorization.

\subsection{Left-right factorization and string theory variables}
The full non-linear field redefinition that relates the fluctuation $\check{e}_{ij}$ to $e_{ij}$
can be written in closed form as\cite{Hull:2009zb,Hohm:2010jy,Michishita:2006dr},
 \be\label{allorder}
  \check{e}_{ij} \ = \ F_{i}{}^{k}(e)\,e_{kj}\;, \qquad
  F \ = \ \left(1-\frac{1}{2}eG^{-1}\right)^{-1}\;,
 \ee
where we used matrix notation.
If the $b$-field is set to zero, this agrees to lowest order with (\ref{fieldred}), but we
note that the non-linear extension differs from the field redefinition proposed in \cite{Bern:1999ji}.

Expanding the double field theory action (\ref{THEAction}) to cubic order in terms of
$e_{ij}$, one arrives at the action that has been derived in \cite{Hull:2009mi}
from closed string field theory \cite{Hohm:2010jy}, and whose quadratic piece we display,
\be
\label{redef-action}
\begin{split}
S^{(2)} \ = \   \hskip-2pt\int d x d\tilde x \,
\Bigl[\, \, {1\over 4} e_{ij} \square e^{ij}  + {1\over 4} (\bar D^j e_{ij})^2
+ {1\over 4} ( D^i e_{ij})^2  - 2 \, d \, D^i \bar D^j e_{ij}
\,-\, 4 \,d \, \square \, d~\Big]\;.
\end{split}
\ee
This action exhibits only consistent left-right index contractions and
a corresponding T-duality property in the following sense.
Instead of the matrices (\ref{MbarM}), the transformation rules are governed in this
background-\textit{dependent} formulation by \cite{Hull:2009mi,Kugo:1992md}
 \be\label{MMbar}
  M \ = \ d^{t}-E\,c^{t}\;, \qquad \bar{M} \ = \ d^{t}+E^{t}\,c^{t}\;,
 \ee
where the constant background $E_{ij}$ rather than ${\cal E}_{ij}$ enters. Similarly,
the calligraphic derivatives (\ref{callder}) are replaced by $D_{i}$ and $\bar{D}_{i}$ depending
only on $E$, while index contractions are done with $G^{ij}$. We now require that the background
transforms under $O(D,D)$ as
 \be\label{Eprime}
  E^{\prime} \ = \ (aE+b)(cE+d)^{-1}\;,
 \ee
which implies for the background metric
 \be\label{Gprime}
  G^{\prime -1} \ = \ \bar{M}^t\,G^{-1}\,\bar{M} \ = \
  M^t\, G^{-1}\,M\;.
 \ee
This is the analogue of  (\ref{fsetmatvmvgn9998}) and thus the indices on $G^{-1}$ can
again be thought of as being either both unbarred or both barred. (Here, we have written
the primed variables in terms of the unprimed ones, because it is this form that will be
used below.)
Moreover, we require that $d$ is invariant, and we prove below that $e_{ij}$ transforms according to
 \be\label{etrans}
  e_{ij} \ = \ M_{i}{}^{k}\,\bar{M}_{j}{}^{l}\,e^{\prime}_{kl}\;.
 \ee
Thus, the left index transforms with $M$ and the right index with $\bar{M}$, and since
the action has only consistent left-right index contractions, it follows that
the action has the T-duality property
 \be\label{Tdualprop}
  S[E^{\prime},e^{\prime},d^{\prime}] \ = \ S[E,e,d]\;.
 \ee

The logic can now be turned around in order to find a prescription that guarantees left-right
factorization to arbitrary orders. If we start from the full non-linear action (\ref{THEAction})
and use the non-linear field redefinition (\ref{allorder}), we can in principle expand the action
in terms of $e_{ij}$ to any desired order. The original action is $O(D,D)$ invariant and, therefore, 
the resulting action has the T-duality property (\ref{Tdualprop}).
Since the action is then written only in terms of objects that transform `covariantly'
with $M$ or $\bar{M}$, it follows that all terms with inconsistent
left-right index contractions cancel out, leaving an action with left-right
factorization.\footnote{This is to be contrasted with expressions that
contain both covariant \textit{and} non-covariant objects, as eq.~(\ref{eecheckrel}) to be
discussed below. In this case, the non-covariant transformation of one term can cancel
against the non-covariant transformation of another term.}

In order to complete the above proof we have to show that $e_{ij}$ transforms according to
(\ref{etrans}) to all orders.
In \cite{Hull:2009mi} this has been verified to lowest order, while the validity of the non-linear
field redefinition (\ref{allorder}) has been confirmed by inspection of the gauge symmetries
in \cite{Hull:2009zb,Hohm:2010jy}. Here we complete the existing literature
by showing that the original $O(D,D)$ transformation (\ref{ETdual}), together
with the form of the field redefinition (\ref{allorder}), indeed implies
the simple transformation rule (\ref{etrans})
to all orders.

We first determine the transformation behavior of $\check{e}$ from (\ref{ETdual}),
 \bea\label{echtrans}
 \begin{split}
  E+\check{e}&\,\rightarrow\, (aE+b+a\check{e})(cE+d+c\check{e})^{-1} \\
  & \ = \ (aE+b+a\check{e})(cE+d)^{-1}\left(1-c\check{e}(cE+d)^{-1}+
  (c\check{e}(cE+d)^{-1})^2\mp\cdots\right) \\
  & \ = \ E^{\prime}+(a-E^{\prime}c)\check{e}(\bar{M}^t)^{-1}\left(1-c\check{e}
  (\bar{M}^t)^{-1}+(c\check{e}(\bar{M}^t)^{-1})^2\mp\cdots\right) \\
  & \ = \ E^{\prime}+(a-E^{\prime}c)\check{e}(\bar{M}^t)^{-1}
  \left(1+c\check{e}(\bar{M}^t)^{-1}\right)^{-1} \\
  & \ = \ E^{\prime}+(a-E^{\prime}c)\check{e}\left(\bar{M}^t+c\check{e}\right)^{-1}
 \;.
\end{split}
\eea
Here we have used the following matrix identity for general $X$ and $Y$,
 \be
  (X+Y)^{-1} \ = \ X^{-1}\left(1-YX^{-1}+(YX^{-1})^2\mp \cdots\right)\;,
 \ee
together with (\ref{MMbar}). Moreover, we have used (\ref{Eprime}) in order to identify
the transformed background.
The final expression in (\ref{echtrans})
can be further simplified using eq.~(4.13) from \cite{Hull:2009mi}, which implies
 \bea
  E^{\prime} \ = \ -M^{-1}(b^t-Ea^t)\;.
 \eea
We also have to use the group properties of $O(D,D)$, which require in particular
 \be
  b^tc+d^ta \ = \ {\bf 1}\;, \qquad a^tc+c^ta  \ = \ 0\;.
 \ee
We then find
 \be
  \begin{split}
  a-E^{\prime}c  \ &= \ a+M^{-1}(b^t-Ea^t)c \\
   \ &= \ a+M^{-1}({\bf 1}-d^ta+Ec^ta)\\
   \ &= \ a+M^{-1}({\bf 1}-Ma)\\
   \ &= \ M^{-1}\;.
  \end{split}
 \ee
Using this in (\ref{echtrans}) we can read off the transformation behavior of $\check{e}$,
 \be\label{echecktrans}
  \check{e}^{\prime} \ = \ M^{-1}\,\check{e}\,\left(\bar{M}^t+c\check{e}\right)^{-1}
  \ = \ M^{-1}\,\check{e}\,(\bar{\cal M}^t)^{-1}\;,
 \ee
where we used in the last equation
 \be
  \bar{M}^t+c\check{e} \ = \ cE+d+c\check{e} \ = \ c{\cal E}+d \ \equiv \
  \bar{\cal M}^t\;.
 \ee
Thus, curiously, the left index of $\check{e}_{ij}$ transforms with $M$, but the right index
with the background-independent $\bar{\cal M}$ defined in (\ref{MbarM}).

Next, we determine from this result the transformation behavior of $e_{ij}$ according to
(\ref{allorder}) in order to verify (\ref{etrans}), which we write here in matrix notation as
 \be\label{emattrans}
  e^{\prime} \ = \ M^{-1}\,e\,(\bar{M}^t)^{-1}\;.
 \ee
To prove this we show that the defining relation (\ref{allorder}) between
$e$ and $\check{e}$, which we write as
 \be\label{eecheckrel}
  \check{e}-\frac{1}{2}e\,G^{-1}\,\check{e} \ = \ e\;,
 \ee
is invariant under (\ref{Gprime}), (\ref{echecktrans}) and (\ref{emattrans}).
Replacing in (\ref{eecheckrel}) all objects by primed objects we get
 \be
  M^{-1}\check{e}(\bar{\cal M}^t)^{-1}-\frac{1}{2}M^{-1}e(\bar{M}^t)^{-1}M^t G^{-1}
  M M^{-1}\check{e} (\bar{\cal M}^t)^{-1} \ = \ M^{-1}\,e\,(\bar{M}^t)^{-1}\;.
 \ee
Multiplying from the left with $M$ and from the right with $\bar{\cal M}^t$ we obtain
 \be\label{noncovstep}
  \check{e}-\frac{1}{2}e (\bar{M}^t)^{-1}M^tG^{-1}\check{e} \ = \
  e(\bar{M}^t)^{-1}\bar{\cal M}^t\;.
 \ee
There are two contributions which are `non-invariant': the terms on the left-hand side
which do not cancel because one matrix is $M$ and the other $\bar{M}$; and the terms on
the right-hand side which do not cancel because one matrix is the background-dependent
$\bar{M}$ and the other the background-independent $\bar{\cal M}$.
We compute the failure of covariance in each case. First,
 \be
 \begin{split}
  (\bar{M}^t)^{-1}\,M^t \ &= \  (\bar{M}^t)^{-1}\left(\bar{M}^t+(M^t-\bar{M}^t)\right)\\
   \ &= \ {\bf 1}+(\bar{M}^t)^{-1}\left(d-cE^t-d-cE\right) \\
   \ &= \ {\bf 1}-2(\bar{M}^t)^{-1}cG\;.
 \end{split}
 \ee
Second,
 \be
  (\bar{M}^t)^{-1}\bar{\cal M}^t \ = \
  (cE+d)^{-1}(cE+d+c\check{e}) \ = \ {\bf 1}+(\bar{M}^t)^{-1}c\check{e}\;.
 \ee
Inserting these into (\ref{noncovstep}) one infers that the non-invariant terms cancel
each other, giving back (\ref{eecheckrel}). This completes the proof that $e$ transforms
according to (\ref{emattrans}).

\section{$GL(D)\times GL(D)$ covariant frame formulation}
In this section we reformulate the above results using the frame formalism developed by Siegel.
We first briefly review the aspects of this formalism that are relevant for our subsequent
analysis; for a more detailed account we refer to the original literature \cite{Siegel:1993th}
or to the recent papers \cite{Hohm:2010pp,Hohm:2010xe}.
Then we turn to the perturbative expansion about a constant background and prove that
the natural frame-like variable can be identified with the
string field theory variable $e_{ij}$ above.

\subsection{Siegels frame formalism}
This formalism is based on the frame field $e_{A}{}^{M}$ that is a $2D\times 2D$ matrix,
 \be\label{Egauge}
   e_{A}{}^{M} \ = \ \begin{pmatrix} e_{ai} &  e_{a}{}^{i} \\ e_{\bar{a}i} & e_{\bar{a}}{}^{i} \end{pmatrix}
   \;.
 \ee
Here, the index $M$ is a fundamental $O(D,D)$ index and splits according to
${}^M = (\,{}_{i}\,,\,{}^{i}\,)$; $A$ is the $GL(D)\times GL(D)$ index
and splits according to $A=(a,\bar{a})$. Thus, the frame field transforms under
global $O(D,D)$ and under local $GL(D)\times GL(D)$
transformations. Moreover, the frame field transforms under a gauge symmetry
with a parameter $\xi^{M}=(\tilde{\xi}_{i},\xi^{i})$ that combines the
usual diffeomorphism parameter $\xi^{i}$ and the $b$-field gauge parameter $\tilde{\xi}_{i}$
into a fundamental $O(D,D)$ vector,
 \be\label{xigauge}
  \delta_{\xi}e_{A}{}^{M} \ = \ \xi^{N}\partial_{N}e_{A}{}^{M}
  +\big(\partial^{M}\xi_{N}-\partial_{N}\xi^{M}\big)e_{A}{}^{N}\;,
 \ee
where $\partial_{M}=(\tilde{\partial}^{i},\partial_{i})$, and indices $M,N,\dots$
are raised and lowered with the $O(D,D)$  invariant metric $\eta_{MN}$.

Next, we define a space-time dependent tangent space metric
from $\eta_{MN}$ using the frame field $e_{A}{}^{M}$,
 \be\label{flatmetric}
  {\cal G}_{AB} \ = \ e_{A}{}^{M}\,e_{B}{}^{N}\,\eta_{MN}\; .
 \ee
This metric will be used to raise and lower flat indices.
In order for the frame field to describe the same degrees of freedom as
the massless sector of closed string theory, it needs to satisfy the
$GL(D)\times GL(D)$ covariant constraint
 \be\label{Goff}
   {\cal G}_{a\bar{b}} \ = \ 0 \quad \Leftrightarrow \quad e_{(a}{}^{i}\,e_{\bar{b})i} \ = \ 0\;.
 \ee

There are various ways to identify the conventional Einstein variables
${\cal E}_{ij}=g_{ij}+b_{ij}$ in this formalism. The most direct way is to
gauge-fix the $GL(D)\times GL(D)$ symmetry by setting the vielbein components
$e_{a}{}^{i}$ and $e_{\bar{a}}{}^{i}$ equal to the unit matrix (assuming
that these vielbeins are invertible),
 \be\label{Egauge2}
   e_{A}{}^{M}
   \ = \ \begin{pmatrix} -{\cal E}_{ai} &   \delta_{a}{}^{i} \\
   {\cal E}_{i\bar{a}} & \delta_{\bar{a}}{}^{i} \end{pmatrix}\;.
 \ee
In here, the constraint (\ref{Goff}) is implemented by parametrizing
the remaining components by a single matrix ${\cal E}_{ij}$.
A $GL(D)\times GL(D)$ covariant definition that does not require a gauge fixing, and which
will be more useful below, is to identify \cite{Hohm:2010pp}
 \be\label{covE}
  {\cal E}_{ij} \ = \ -e_{i}{}^{a}\,e_{aj} \ = \ e_{\bar{a}i}\,e_{j}{}^{\bar a}\;,
 \ee
where $e_{i}{}^{a}$ is the inverse of $e_{a}{}^{i}$, and $e_{i}{}^{\bar a}$ is
the inverse of $e_{\bar a}{}^{i}$. We stress that the vielbeins
in (\ref{covE}) are independent, and the flat indices are not raised or
lowered by means of an invariant tensor (`two-vierbein formalism' \cite{Siegel2}).
For the gauge choice (\ref{Egauge2}) the definition (\ref{covE}) coincides
with the previous definition of ${\cal E}_{ij}$, but (\ref{covE}) is more general
in that it holds for arbitrary gauge choices.
The metric $g_{ij}$ can then be obtained from the tangent space metric
(\ref{flatmetric}) according to
 \be\label{metricrel}
  g_{ij} \ = \ -\frac{1}{2}e_{i}{}^{a}\,e_{j}{}^{b}\,{\cal G}_{ab}
  \ = \ \frac{1}{2}e_{i}{}^{\bar a}\,e_{j}{}^{\bar b}\,{\cal G}_{\bar a\bar b}\;.
 \ee

Next, we mention that one can introduce connections $\omega_{A}$ for the $GL(D)\times GL(D)$
tangent space symmetry in order to construct covariant derivatives
$\nabla_{A}=e_{A}+\omega_{A}$, where $e_{A}=e_{A}{}^{M}\partial_{M}$ is
the ordinary (but `flattened') derivative. Without repeating details here,
we record that the calligraphic derivatives of ${\cal E}$ can be identified
with $GL(D)\times GL(D)$ covariant derivatives of $e_{A}{}^{M}$ as follows
\cite{Hohm:2010xe},
 \be\label{covderE}
 \begin{split}
  &e_{a}{}^{i}\,e_{b}{}^{j}\,e_{\bar c}{}^{k}\,{\cal D}_{i}{\cal E}_{jk} \ = \
  e_{b}{}^{M}\nabla_{a}e_{{\bar c}M} \ = \ -e_{\bar c}{}^{M}\nabla_{a}e_{bM}\;, \\
  &e_{\bar a}{}^{i}\,e_{b}{}^{j}\,e_{\bar c}{}^{k}\,\bar{\cal D}_{i}{\cal E}_{jk} \ = \
  e_{b}{}^{M}\nabla_{\bar a}e_{{\bar c}M} \ = \ -e_{\bar c}{}^{M}\nabla_{\bar a}e_{bM}
  \;.
 \end{split}
 \ee
Even though here we have used covariant derivatives in order to make the full tangent space
symmetry manifest, we note that the connections in (\ref{covderE}) actually drop out as a consequence
of (\ref{Goff}), which will be used below.
The double field theory action is then equivalent to
\cite{Hohm:2010xe}
 \be\label{frameaction}
 \begin{split}
  S \ = \ \int dx d\tilde{x}\,e^{-2d}\,\Big[\;&{\cal G}^{ab}\,{\cal G}^{\bar c\bar d}\,
  \Big(-\frac{1}{2}e_{a}{}^{M}\nabla^{c}e_{\bar c M}\,e_{b}{}^{N}\nabla_{c}e_{\bar d N}
  +\frac{1}{2}e_{\bar c}{}^{M}\nabla_{a}e_{c M}\,e_{\bar d}{}^{N}\nabla^{c}e_{bN} \\
  &-\frac{1}{2}e_{a}{}^{M}\nabla^{\bar a}e_{\bar c M}\; e_{b}{}^{N}\nabla_{\bar d}e_{\bar a N}
  -e_{a}{}^{M}\nabla_{\bar c}e_{\bar d M}\,\nabla_{b}d+e_{\bar c}{}^{M}\nabla_{a}e_{b M}\,\nabla_{\bar d}d
  \Big) \\
  &-2\nabla^{a}d\,\nabla_{a}d\;\Big]
   \;.
 \end{split}
 \ee
Using the relations (\ref{metricrel}) and (\ref{covderE}) between `tangent
space' and `world' objects, it can be easily seen, upon converting indices with
$e_{a}{}^{i}$ and $e_{\bar a}{}^{i}$, that this action is
equal to (\ref{THEAction}), up to an irrelevant overall factor.

\subsection{Perturbation theory in terms of frame-like variables}
We next discuss the linearization around a constant background \cite{Siegel:1993th},
 \be\label{ehexp}
  e_{A}{}^{M} \ = \ \langle e_{A}{}^{M}\rangle -h_{A}{}^{B}\langle e_{B}{}^{M}\rangle\;,
 \ee
where we introduced a fluctuation $h_{AB}$ with flat indices (that are raised and lowered
with the background tangent space metric $\vev{{\cal G}_{AB}}$).
This expansion is meant to be exact, i.e., all higher powers in $h$
in the full theory will originate from taking the inverse of this expression.

Let us first examine the
$GL(D)\times GL(D)$ gauge symmetries, whose infinitesimal form we define to be
$\delta_{\Lambda}e_{A}{}^{M} = \Lambda_{A}{}^{B}e_{B}{}^{M}$, with the only non-trivial parameters
$\Lambda_{a}{}^{b}$ and $\Lambda_{\bar a}{}^{\bar b}$.
As for the field, we expand also the gauge parameter into a background and a first-order part,
 \be
  \Lambda_{A}{}^{B} \ = \ \bar{\Lambda}_{A}{}^{B}-\epsilon_{A}{}^{B}\;,
 \ee
where $\bar{\Lambda}$ is constant and thus represents a \textit{global}
$GL(D)\times GL(D)$ symmetry.\footnote{One may compare this with a similar splitting in
the conventional formulation of gravity. Here, after expansion about a flat background, the
diffeomorphism symmetry gives rise to two types of symmetries: global Poincar\'e transformations
and local gauge transformations $\delta h_{ij}=\partial_{i}\xi_{j}+\partial_{j}\xi_{i}$.}
Acting on (\ref{ehexp}),
 \be
  \delta_{\Lambda}\big(\langle e_{A}{}^{M}\rangle -h_{A}{}^{B}\langle e_{B}{}^{M}\rangle\big)
   \ = \ \big(\bar{\Lambda}_{A}{}^{B}-\epsilon_{A}{}^{B}\big)\big(\langle e_{B}{}^{M}\rangle
   -h_{B}{}^{C}\langle e_{C}{}^{M}\rangle\big)\;,
  \ee
we read off by comparing the orders
 \be\label{expgauge}
  \begin{split}
   \delta_{\Lambda}\langle e_{A}{}^{M}\rangle  \ &= \ \bar{\Lambda}_{A}{}^{B}\langle e_{B}{}^{M}\rangle\;,
   \\
   \delta_{\Lambda}h_{AB} \ &= \ \epsilon_{AB}-\epsilon_{A}{}^{C}h_{CB}
   +\bar{\Lambda}_{A}{}^{C}h_{CB}+
   \bar{\Lambda}_{B}{}^{C}h_{AC}\;,
  \end{split}
 \ee
where we used
 \be
    \delta_{\Lambda}h_{AB} \ = \
   \delta_{\Lambda}h_{A}{}^{C}\,\langle {\cal G}_{CB}\rangle
   +h_{A}{}^{C}\delta_{\Lambda}\langle {\cal G}_{CB}\rangle\;, \qquad
   \delta_{\Lambda}\langle{\cal G}_{AB} \rangle \ = \ \bar{\Lambda}_{AB}+  \bar{\Lambda}_{BA}
  \;.
 \ee
We infer that to lowest order the fluctuations are subject to a shift symmetry,
$\delta_{\epsilon} h_{ab}  =  \epsilon_{ab}$ and
$\delta_{\epsilon} h_{\bar{a}\bar{b}} =  \epsilon_{\bar{a}\bar{b}}$, and so it is natural
to impose the gauge fixing condition
 \be\label{firstgauge}
  h_{ab} \ = \ h_{\bar{a}\bar{b}} \ = \ 0\;.
 \ee
This is also meant to be an exact gauge fixing condition. The constraint
(\ref{Goff}) then implies
 \be\label{offrel}
  \begin{split}
   0 \ &= \  e_{a}{}^{M}\,e_{\bar b}{}^{N}\,\eta_{MN} \ = \
   \big(\vev{e_{a}{}^{M}}-h_{a}{}^{\bar c}\vev{e_{\bar c}{}^{M}}\big)
   \big(\vev{e_{\bar b}{}^{N}}-h_{\bar b}{}^{d}\vev{e_{d}{}^{N}}\big)\eta_{MN} \\
   \ &= \ \langle {\cal G}_{a\bar{b}}\rangle - h_{a\bar{b}}  - h_{\bar{b}a}
   \qquad \Rightarrow\qquad h_{a\bar{b}} +h_{\bar{b}a} \ = \ 0\;.
 \end{split}
 \ee
Here we used that the background tangent space metric satisfies the constraint
 \be\label{backoff}
  0 \ = \ \vev{{\cal G}_{a\bar{b}}} \ = \ \vev{e_{a}{}^{M}}\vev{e_{\bar b}{}^{N}}\eta_{MN}\;.
 \ee
We infer from (\ref{firstgauge}) and (\ref{offrel}) that $h_{a\bar{b}}$ is the only independent component.

After having fixed the gauge symmetries spanned by $\epsilon_{AB}$, let us briefly turn to
the background $GL(D)\times GL(D)$ transformations parametrized by $\bar{\Lambda}_{A}{}^{B}$.
In order to make contact with the formalism in sec.~2, we fix this global symmetry, in analogy to (\ref{Egauge2}), by setting the background frame field equal to
 \be\label{backgauge}
   \vev{e_{A}{}^{M}} \ = \ \begin{pmatrix} -E_{ai} &  \delta_{a}{}^{i} \\ E_{i\bar{a}} & \delta_{\bar{a}}{}^{i}
   \end{pmatrix}\;.
 \ee
We stress that the complete gauge fixing of the $GL(D)\times GL(D)$ symmetry consisting of
(\ref{firstgauge}) and (\ref{backgauge}) is inequivalent to the gauge fixing in (\ref{Egauge2}),
because here $e_{a}{}^{i}=\delta_{a}{}^{i}$, etc., holds only in the background.

Before imposing the condition (\ref{backgauge}), the fluctuations $h_{AB}$ are
inert under global $O(D,D)$ transformations, because they carry only flat indices.
After the gauge fixing, however, compensating background
$GL(D)\times GL(D)$ transformations are required, which will lead to non-trivial
$O(D,D)$ transformations of $h_{AB}$. We write the finite form of the background 
$GL(D)\times GL(D)$ transformations 
generated by $\bar{\Lambda}$ in (\ref{expgauge}) as
 \be
  \langle e_{a}{}^{M\prime}\rangle \ = \ ( M^{-1})_{a}{}^{b}\,\langle e_{b}{}^{M}\rangle\;,
  \qquad
  \langle e_{\bar a}{}^{M\prime}\rangle \ = \ ( \bar{M}^{-1})_{\bar a}{}^{\bar b}\,\langle e_{\bar b}{}^{M}\rangle\;,
 \ee
and thus for the independent fluctuation as
 \be\label{hflattrans}
   h_{a\bar{b}}^{\prime} \ = \  ( M^{-1})_{a}{}^{c}\,(\bar{M}^{-1})_{\bar b}{}^{\bar d}\,h_{c\bar{d}}\;.
 \ee
If we require now that the matrices $M$ and $\bar M$ are determined in terms of the $O(D,D)$
matrix in such a way that the form of the background frame field (\ref{backgauge}) be preserved,
we find (in complete analogy to the analysis of sec.~4.1 in \cite{Hohm:2010xe})
 \be\label{Magain}
  M \ = \ d^t-E c^t\;, \qquad
  \bar{M} \ = \ d^{t}+E^{t}c^{t}\;,
 \ee
while $E_{ij}$ transforms according to (\ref{Eprime}). These matrices coincide with (\ref{MMbar}),
and thus we recovered the formalism of sec.~2, in which unbarred indices transform with
$M$ and barred indices transform with $\bar M$. More precisely, via the trivial background
vielbeins $\vev{e_{a}{}^{i}}=\delta_{a}{}^{i}$ and $\vev{e_{\bar a}{}^{i}}=\delta_{\bar a}{}^{i}$ one can
identify indices $i,j,\ldots$ with flat unbarred or barred indices that transform with (\ref{Magain}).
From (\ref{hflattrans}) we conclude that $h_{a\bar b}$ transforms in the same way as $e_{ij}$
under $O(D,D)$. Indeed, we will prove below that these two variables are identical for
the background choice (\ref{backgauge}).
In the following we present all results more generally, i.e., not assuming that the background takes
the specific form (\ref{backgauge}), unless stated differently. The field $e_{ij}$ is then
related to $h_{a\bar b}$ via
 \be\label{framestring}
  e_{ij} \ = \ \vev{e_{i}{}^{a}}\,\vev{e_{j}{}^{\bar b}}\,h_{a\bar b}\;.
 \ee
In particular, this variable transforms under $O(D,D)$ according to (\ref{etrans}) in general.

Let us now turn to the gauge symmetries (\ref{xigauge}) parametrized by $\xi^{M}$.
We assume from now on that the background is inert under gauge transformations,
and thus we obtain for the gauge transformation of the fluctuation
 \be
   \delta_{\xi}h_{A}{}^{B}\,\langle e_{B}{}^{M}\rangle \ = \
   \xi^{N}\partial_{N}h_{A}{}^{B}\,\langle e_{B}{}^{M}\rangle-\big(\partial^{M}\xi_{N}
   -\partial_{N}\xi^{M}\big)\big(\langle e_{A}{}^{N}\rangle -h_{A}{}^{B}\langle e_{B}{}^{N}\rangle\big)\;.
  \ee
Multiplying with $\langle e_{CM}\rangle$, relabeling indices and rearranging terms we find
 \be\label{delxih}
  \delta_{\xi}h_{AB} \ = \  D_{A}\xi_{B}-D_{B}\xi_{A}+
  \xi^{N}\partial_{N}h_{AB}+\big(D_{B}\xi^{C}-D^{C}\xi_{B}\big)h_{AC}\;,
 \ee
where we introduced gauge parameters and derivatives whose indices have been flattened
with the background frame field,
 \be
  D_{A} \ = \ \vev{e_{A}{}^{M}}\partial_{M}\;, \qquad
  \xi_{A} \ = \ \vev{e_{A}{}^{M}}\xi_{M}\;.
 \ee
The derivatives $D_{A}$ are related to the derivatives in sec.~2 through
 \be
  D_{i} \ = \ \vev{e_{i}{}^{a}}\,D_{a}\;, \qquad
  \bar{D}_{i} \ = \ \vev{e_{i}{}^{\bar a}}\,D_{\bar a}\;,
 \ee
i.e., they agree for the choice (\ref{backgauge}) of the background.
Using that the background metric $G_{ij}$ is related to the tangent space metric via
 \be\label{Gformula}
  G_{ij} \ = \ -\frac{1}{2}\vev{e_{i}{}^{a}}\,\vev{e_{j}{}^{b}}\,\vev{{\cal G}_{ab}}
  \ = \ \frac{1}{2}\vev{e_{i}{}^{\bar a}}\,\vev{e_{j}{}^{\bar b}}\,\vev{{\cal G}_{\bar a\bar b}}\;,
 \ee
we note that the constraint (\ref{strongconstr}) translates into 
 \be\label{frameconstr}
  D^{a} D_{a} + D^{\bar a} D_{\bar{a}} \ = \ 0\;,
 \ee 
when acting on fields and their products.  

Next, we inspect the gauge variation of the gauge-fixed components,
 \be
 \begin{split}
  \delta_{\xi}h_{ab} \  &= \  D_{a}\xi_{b}-D_{b}\xi_{a}+\big(D_{b}\xi^{\bar c}
  -D^{\bar c}\xi_{b}\big)h_{a\bar{c}}\;, \\[1ex]
  \delta_{\xi}h_{\bar a\bar b} \  &= \ D_{\bar a}\xi_{\bar b}
  -D_{\bar b}\xi_{\bar a}
  +\big(D_{\bar b}\xi^{c}-D^{c}\xi_{\bar b}\big)h_{\bar a c}\;,
 \end{split}
 \ee
where we made use of the gauge fixing condition (\ref{firstgauge}).
Thus, the gauge condition is not invariant under
$\xi^{M}$ transformations. This requires compensating gauge transformations parametrized by
$\epsilon_{AB}$. Choosing $\epsilon_{ab}$ to be
 \be\label{epsshell}
  \epsilon_{ab} \ = \ D_{b}\xi_{a}-D_{a}\xi_{b}-\big(D_{b}\xi^{\bar c}-
  D^{\bar{c}}\xi_{b}\big)h_{a\bar{c}}\;,
 \ee
restores the gauge condition $h_{ab}=0$.
With this form of the compensating gauge transformation we can compute the
complete gauge variation of $h_{a\bar b}$ from (\ref{expgauge}) and (\ref{delxih}),
 \be\label{delh3}
 \begin{split}
  \delta_{\xi}h_{a\bar b} \ = \ &D_{a}\xi_{\bar b}-D_{\bar b}\xi_{a}
  +\big(\xi^{c} D_{c}+\xi^{\bar c}D_{\bar c}\big)h_{a\bar b}
  +\big( D_{\bar b}\xi^{\bar c}-D^{\bar c}\xi_{\bar b} \big)h_{a\bar c}
  -\epsilon_{a}{}^{c}\,h_{c\bar b}\;. \\
  \ = \ &D_{a}\xi_{\bar b}-D_{\bar b}\xi_{a} \\
  &+\big(\xi^{c} D_{c}+\xi^{\bar c}D_{\bar c}\big)h_{a\bar b}
  +\big( D_{\bar b}\xi^{\bar c}-D^{\bar c}\xi_{\bar b} \big)h_{a\bar c}
  +\big(D_{a}\xi^{c}-D^{c}\xi_{a}\big)h_{c\bar b} \\
  &+h_{a\bar d}\big(D^{c}\xi^{\bar d}-D^{\bar d}\xi^{c}\big)h_{c\bar b}\;,
 \end{split}
 \ee
where we replaced $\epsilon$ in the second equation by (\ref{epsshell}).

In order to compare this result with the gauge transformation of the string field theory
variable (\ref{framestring}) as determined in \cite{Hull:2009zb}, we introduce gauge parameters
according to
 \be
  \lambda_{i} \ = \ -\vev{e_{i}{}^{a}}\,\xi_{a}\;, \qquad
  \bar{\lambda}_{i} \ = \ \vev{e_{i}{}^{\bar a}}\,\xi_{\bar a}\;,
 \ee
and we assume that indices are contracted with the background metric $G_{ij}$ in (\ref{Gformula}). 
We then obtain
 \be\label{delstring}
  \begin{split}
   \delta_{\lambda}e_{ij} \ = \ &D_{i}\bar{\lambda}_{j}+\bar{D}_{j}\lambda_{i}\\
   &+\frac{1}{2}\big(\lambda\cdot D+\bar{\lambda}\cdot\bar{D}\big)e_{ij}
   +\frac{1}{2}\big(\bar{D}_{j}\bar{\lambda}^{k}-\bar{D}^{k}\bar{\lambda}_{j}\big)e_{ik}
   +\frac{1}{2}\big(D_{i}\lambda^k -D^k \lambda_i\big)e_{kj}\\
   &-\frac{1}{4}e_{ik}\big(D^l \bar{\lambda}^{k}+\bar{D}^{k}\lambda^{l}\big)e_{lj}\;,
  \end{split}
 \ee
which agrees precisely with eq.~(2.20) in \cite{Hull:2009zb}.

So far we have seen that $h_{a\bar{b}}$ gives rise, via (\ref{framestring}), 
to a variable that transforms under $O(D,D)$
with the matrices $M$ and $\bar M$ and that transforms under gauge transformations 
as required by the exact result (\ref{delstring}). This shows that
$h_{a\bar b}$ can be identified with the string field theory variable $e_{ij}$ according
to (\ref{framestring}). In fact, while it is possible to perform field redefinitions that
preserve the left-right index structure,
e.g.
 \be\label{redefinition} 
  h_{a\bar{b}}\;\rightarrow\; h_{a\bar b}+\alpha \,h_{a}{}^{\bar c}\,h_{d\bar c}\,h^{d}{}_{\bar b}
  +{\cal O}(h^5)\;, 
 \ee
this would induce higher order terms in the gauge transformation
(\ref{delh3}) and thus be inconsistent with the form (\ref{delstring}).

We close this section by verifying that $h_{a\bar b}$ is related to the Einstein variable
$\check{e}$ according to the field redefinition (\ref{allorder}), which provides a direct
proof for the above conclusion.
In order to simplify the notation, we assume that the background takes the specific form (\ref{backgauge}),
such that we can identify $e_{ij}$ and $h_{a\bar b}$ directly.
We use matrix notation and denote the matrix with components $E_{ij}$ by $E$, the matrix
with components $e_{ai}$ by $e_{*}$, the matrix with components $e_{a}{}^{i}$ by $e^{*}$ and
the matrix with components $h_{a}{}^{\bar b}$ by $h$.
The expansions
 \be
  e_{ai} \ = \  -E_{ai} - h_{a}{}^{\bar b}\,E_{i\bar b}\;, \qquad
  e_{a}{}^{i} \ = \ \delta_{a}{}^{i}-h_{a}{}^{\bar b\,}\delta_{\bar b}{}^{i}\;,
 \ee
then read
 \be\label{matrixexp}
  e_{*} \ = \ -E-h\,E^t\;, \qquad
  e^{*} \ = \ 1-h\;,
 \ee
while ${\cal E}$ becomes according to (\ref{covE})
 \be
  {\cal E} \ =  \ -\left(e^{*}\right)^{-1}\, e_{*}\;.
 \ee
Inserting here the expansions (\ref{matrixexp}), we obtain
 \be
 \begin{split}
  {\cal E} \ &= \ \left(1-h\right)^{-1}\left(E+h\,E^{t}\right) \\
  \ &= \ \left(1+h+h^2+h^3+\cdots\right) \left(E+h\,E^t\right) \\
  \ &= \ E+h\left(E+E^t\right) +h^2\left(E+E^t\right)+h^3\left(E+E^t\right)+\cdots \\
  \ &= \ E+2hG+2h^2G+2h^3G+\cdots\;,
 \end{split}
 \ee
where we used $G=\tfrac{1}{2}(E+E^t)$.
Comparison with ${\cal E}=E+\check{e}$ then implies 
 \be\label{step965}
  \check{e} \ = \ 2\left(1+h+h^2+\cdots\right)hG \ = \
  2\left(1-h\right)^{-1}hG\;.
 \ee
The identification (\ref{framestring}) relates $e_{ij}$ to 
$h_{a}{}^{\bar c}\langle {\cal G}_{\bar c\bar b}\rangle$, which yields with (\ref{Gformula})
 \be
  e \ = \ 2\,h\,G \;.
 \ee
Inserting this into (\ref{step965}) finally implies
 \be
  \check{e} \ = \ \left(1-\frac{1}{2}eG^{-1}\right)^{-1}e\;,
 \ee
exactly as required by (\ref{allorder}). Thus, we recovered the field redefinition between
Einstein and string theory variables from the frame formalism.

\subsection{The action in terms of frame-like variables}
We turn now to the expansion of the action (\ref{frameaction}) in terms of the frame-like variable
$h_{a\bar b}$. Moreover, we introduce $e^{-d}$ as a fundamental variable with a corresponding
fluctuation,
 \be
  \Phi^2 \ \equiv \ e^{-2d} \; \qquad  \Phi \ = \ 1+\varphi\;.
 \ee
This implies that the new field $\varphi$ is related to $d$ via
 \be\label{didenti}
  \varphi \ = \ -d+\frac{1}{2}d^2+\cdots\;.
 \ee
 The Lagrangian corresponding to (\ref{frameaction}) then reads in terms of
 $\Phi$,\footnote{Using identities like (\ref{covderE}) and (\ref{frameconstr}) 
the action could be rewritten in various ways,
for instance such that it becomes manifest, at the cost of extra terms,
that the barred and unbarred indices enter on the same footing.}
 \be\label{newframeaction}
 \begin{split}
  {\cal L} \ = \ &\,-\frac{1}{2}\Phi^2\,{\cal G}^{ab}\,{\cal G}^{\bar c\bar d}\,
  \Big(\,{\cal G}^{cd}\,e_{a}{}^{M}\nabla_{c}e_{\bar c M}\,e_{b}{}^{N}\nabla^{}_{d}e_{\bar d N}
  -{\cal G}^{cd} e_{\bar c}{}^{M}\nabla_{a}e_{c M}\,e_{\bar d}{}^{N}\nabla_{d}e_{bN}
  \\
  &\qquad\qquad\qquad\quad \;\;
  +{\cal G}^{\bar a\bar b}e_{a}{}^{M}\nabla_{\bar a}e_{\bar c M}\; e_{b}{}^{N}\nabla_{\bar d}e_{\bar b N} \,\Big)\\
  &\,\;+\Phi\, {\cal G}^{ab}\,{\cal G}^{\bar c\bar d}\,
  \Big(e_{a}{}^{M}\nabla^{}_{\bar c}e_{\bar d M}\,\nabla_{b}\Phi
  -e_{\bar c}{}^{M}\nabla_{a}e_{b M}\,\nabla_{\bar d}\Phi
  \Big)
  -2\,{\cal G}^{ab}\,\nabla_{a}\Phi\,\nabla_{b}\Phi
   \;.
 \end{split}
 \ee
Next, we work out the various expressions in here, using that the connections inside the
covariant derivatives drop out \cite{Hohm:2010xe}.
For instance, we find
 \be\label{hexp1}
 \begin{split}
  e_{a}{}^{M}\nabla_{b}e_{\bar c M} \ &= \ \big(\vev{e_{a}{}^{M}}-h_{a}{}^{\bar b}\vev{e_{\bar b}{}^{M}}\big)
  \big(\vev{e_{b}{}^{N}}-h_{b}{}^{\bar d}\vev{e_{\bar d}{}^{N}}\big)\partial_{N}\big(-h_{\bar c}{}^{d}
  \vev{e_{d M}}\big) \\
  \ &= \ D_{b}h_{a\bar c}-h_{b\bar d}D^{\bar d}h_{a\bar c}\;,
 \end{split}
 \ee
where the constraint (\ref{backoff}) has been used.
Similarly,
  \be\label{hexp2}
   \begin{split}
    e_{\bar c}{}^{M}\nabla_{a}e_{bM} \ &= \ -D_{a}h_{b\bar c}+h_{a\bar d}D^{\bar d}h_{b\bar c} \;, \quad
    e_{a}{}^{M}\nabla_{\bar b}e_{\bar c M} \ = \ D_{\bar b}h_{a\bar c}+h_{d\bar b}D^{d}h_{a\bar c}\;, \\
    \nabla_{a}\Phi \ &= \ D_{a}\varphi-h_{a\bar b}D^{\bar b}\varphi\;. \qquad\qquad \qquad
    \nabla_{\bar a}\Phi \ = \ D_{\bar a}\varphi+h_{b\bar a}D^{b}\varphi\;.
    \end{split}
  \ee
Inserting this into (\ref{newframeaction}), we obtain an action which manifestly preserves the
left-right structure. Moreover, the choice of field basis employed here is such that non-polynomial
couplings originate exclusively from taking the inverses
of ${\cal G}_{AB}$. This tangent space metric is given in terms of the fluctuation by
 \be
  {\cal G}_{ab} \ = \ \vev{{\cal G}_{ab}}+h_{a}{}^{\bar c}\,h_{b\bar c}\;, \qquad
  {\cal G}_{\bar a\bar b} \ = \ \vev{{\cal G}_{\bar a\bar b}}+h^{c}{}_{\bar a}\,h_{c\bar b}\;,
 \ee
which are exact relations.
Its inverse can be written in closed form using matrix notation where ${\cal G}$ is the matrix
with components ${\cal G}_{ab}$, $\bar{\cal G}$ is the matrix with components ${\cal G}_{\bar a\bar b}$
and, as above, $h$ is the matrix with components $h_{a}{}^{\bar b}$,
 \be
  {\cal G}^{-1} \ = \ \vev{{\cal G}}^{-1}
  \sum_{n=0}^{\infty} (-1)^n \big(h\, \vev{{\bar{{\cal G}}}}\,h^{t}\,\vev{{\cal G}}^{-1}\big)^n \;,
 \ee
and similarly for ${\cal G}^{\bar a\bar b}$.  Restoring explicit index notation, this reads
 \be\label{InvG}
 \begin{split}
  {\cal G}^{ab} \ &= \ \vev{{\cal G}^{ab}}-h^{a}{}_{\bar c}\,h^{b\bar c}+
  h^{a}{}_{\bar c}\,h^{d\bar c}\,h_{d\bar e}\,h^{b\bar e}\mp\cdots\;, \\
  {\cal G}^{\bar a\bar b} \ &= \ \vev{{\cal G}^{\bar a\bar b}}-h^{c\bar a}\,h_{c}{}^{\bar b}+
  h^{c\bar a}\,h_{c}{}^{\bar d}\,h_{e\bar d}\,h^{e\bar b}\mp \cdots\;.
 \end{split}
 \ee
Using (\ref{hexp1}), (\ref{hexp2}) and (\ref{InvG}) it is, in principle, straightforward to read
off the $n$-point couplings from (\ref{newframeaction}) to any desired order. By 
virtue of using the frame-like variable $h_{a\bar b}$, the left-right factorization is manifest
without further field redefinitions.

As an illustration, we display the quadratic and cubic Lagrangians and verify their equivalence
with the results in the literature. The free Lagrangian reads
 \be\label{2point}
  {\cal L}^{(2)} \ = \ -\frac{1}{2}D^{a}h^{b\bar c}\,D_{a}h_{b\bar c}+\frac{1}{2}D^{a}h^{b\bar c}\,
  D_{b}h_{a\bar c}-\frac{1}{2}D^{\bar a}h^{b\bar c}\,D_{\bar c}h_{b\bar a}
  +2D_{\bar b}h^{a\bar b}\,D_{a}\varphi-2D^{a}\varphi D_{a}\varphi\,.
 \ee
This is equivalent to (\ref{redef-action}), using the identifications (\ref{Gformula}) and (\ref{didenti}).
The cubic Lagrangian reads
 \be
 \begin{split}
  {\cal L}^{(3)} \ = \ \, &h_{a\bar b}\left(D^{a}h^{c\bar d}\,D^{\bar b}h_{c\bar d}-D^{c}h^{a\bar d}\,
  D^{\bar b}h_{c\bar d}-D^{a}h_{c\bar d}\,D^{\bar d}h^{c\bar b}\right) \\
  &+\varphi\left(D^{a}h^{b\bar c}\,D_{b}h_{a\bar c}-D^{a}h^{b\bar c}\,D_{a}h_{b\bar c}
  -D^{\bar a}h^{b\bar c}\,D_{\bar c}h_{b\bar a}+D_{\bar b}h^{a\bar b}\,D_{a}\varphi
  +D_{a}h^{a\bar b}\,D_{\bar b}\varphi\right)\\
  &+h_{a\bar b}\left( D^{a}h^{c\bar b}\,D_{c}\varphi+D_{c}h^{c\bar b}\,D^{a}\varphi
  -D_{\bar c}h^{a\bar c}\,D^{\bar b}\varphi-D^{\bar b}h^{a\bar c}\,D_{\bar c}\varphi
  +4\,D^{a}\varphi \,D^{\bar b}\varphi\right)\;.
 \end{split}
 \ee
Next, we compare this with the cubic action in \cite{Hull:2009mi}. The $D\varphi$  terms in the second line can be rewritten as $D(\varphi^2)$ and then partially
integrated. Moreover, we partially integrate in the third line in order to move first derivatives
away from $\varphi$, after which the Lagrangian is equivalent to
 \be
 \begin{split}
  {\cal L}^{(3)\,\prime} \ = \ \, &h_{a\bar b}\left(D^{a}h^{c\bar d}\,D^{\bar b}h_{c\bar d}-D^{c}h^{a\bar d}\,
  D^{\bar b}h_{c\bar d}-D^{a}h_{c\bar d}\,D^{\bar d}h^{c\bar b}\right) \\
  &-\varphi\left(D^{a}h^{b\bar c}\,D_{a}h_{b\bar c}+\big(D^{a}h_{a\bar b}\big)^2-
  \big(D^{\bar b}h_{a\bar b}\big)^2
  +2h_{a\bar b}\big(D^{a}D_{c}h^{c\bar b}-D^{\bar b}D_{\bar c}h^{a\bar c}\big)\right)\\
  &-\varphi^2D_{a}D_{\bar b}h^{a\bar b}+4\,h_{a\bar b}\,D^{a}\varphi \, D^{\bar b}\varphi\;.
 \end{split}
 \ee
Now we have to
rewrite this in terms of $d$, using the non-linear relation (\ref{didenti}). Specifically,
performing this field redefinition, the
quadratic Lagrangian (\ref{2point}) gives a contribution to the cubic couplings, which
finally read, up to total derivatives,
 \be
 \begin{split}
  {\cal L}^{(3)\,\prime\prime} \ = \ \, &h_{a\bar b}\left(D^{a}h^{c\bar d}\,D^{\bar b}h_{c\bar d}-D^{c}h^{a\bar d}\,
  D^{\bar b}h_{c\bar d}-D^{a}h_{c\bar d}\,D^{\bar d}h^{c\bar b}\right) \\
  &+d\left(D^{a}h^{b\bar c}\,D_{a}h_{b\bar c}+\big(D^{a}h_{a\bar b}\big)^2-
  \big(D^{\bar b}h_{a\bar b}\big)^2
  +2h_{a\bar b}\big(D^{a}D_{c}h^{c\bar b}-D^{\bar b}D_{\bar c}h^{a\bar c}\big)\right)\\
  &-4h_{a\bar b}\,dD^{a}D^{\bar b}d-2d^2\square d\;,
 \end{split}
 \ee
where $\square =D^{a}D_{a}=-D^{\bar a}D_{\bar a}$.  
Using the identifications (\ref{Gformula}) and (\ref{didenti}), this coincides with the
cubic couplings given in eq.~(3.25) of  \cite{Hull:2009mi}.

We close this section with a brief discussion of possible gauge-fixing terms to be added to the
action, which is necessary in order to obtain an invertible propagator and thus to derive the
Feynman rules. For the free theory, a natural choice of gauge conditions is given by  \cite{Siegel:1993th}
 \be\label{gfcond}
  f_{a} \ := \ D^{\bar b}h_{a\bar b}-D_{a}\varphi \ = \ 0\;, \qquad
  f_{\bar a} \ := \ -D^{b}h_{b\bar a}-D_{\bar a}\varphi \ = \ 0\;,
 \ee
which to lowest order
reduces to the usual de Donder gauge-fixing condition if the $b$-field, the (scalar) dilaton
$\phi$ and the tilde derivatives $\tilde{\partial}$ are set to zero.  
The gauge-fixed quadratic action then reads \cite{Siegel:1993th}
 \be
  {\cal L}^{(2)}_{\rm g.f.} \ = \ {\cal L}^{(2)}+\frac{1}{2}\left(f^{a}f_{a}-f^{\bar a}f_{\bar a}\right)
  \ = \ \frac{1}{2}h^{a\bar b}\square h_{a\bar b} +\varphi\square \varphi\;.
 \ee
The gauge conditions (\ref{gfcond}) can be taken to be exact, in which case they translate via
the field redefinition (\ref{allorder}) into a non-linear generalization of the de Donder gauge-fixing
condition for the usual metric fluctuation,
but any non-linear extension of (\ref{gfcond}) that is $GL(D)\times GL(D)$ covariant
would also be consistent with the required left-right factorization to all orders.

\section{Summary and Outlook}
In this note we have discussed the perturbative expansion of the double field theory 
formulation of the low-energy gravity action of closed string theory around a flat background.
When expressed in the field basis natural for string field theory, this Lagrangian 
exhibits a left-right factorization to all orders in perturbation theory by virtue of 
its T-duality invariance. Moreover, we established the precise relation between 
the perturbation theory in this formulation and in the frame-like formalism of Siegel.
In particular, we showed that the string field theory variable coincides precisely
with a frame-like fluctuation in Siegel's formalism. This allows for significant 
technical simplifications when expanding the action since the field redefinition   
relating Einstein to string theory variables is already encoded in the frame-like 
fluctuation. Finally, this relationship might be useful for a more geometrical 
understanding of string field theory.

The manifest left-right factorization at the level of the Lagrangian discussed here 
elevates to a corresponding factorization of the Feynman rules: each Feynman graph 
contributing to an amplitude will factorize into left- and
right-handed parts. This formulation exhibits, therefore, properties inherited from string theory and 
provides the first step towards the goal to render, for instance, the KLT relations more manifest.
The accomplishment of this program requires, however, further work because  
the left- and right-handed factors of the action in terms of $h_{a\bar b}$, which is 
still non-polynomial, remain to be matched with Yang-Mills theory.
In order to achieve this, one procedure could be, as in \cite{Bern:2010yg},
to bring both the gravity and Yang-Mills action to cubic order upon introducing
auxiliary fields. Such a set of auxiliary fields is not unique, but one might
hope that the present formulation will eventually suggest a natural choice. 
We should stress, however, that the field basis discussed here may still be redefined 
in a way consistent with the factorization property, as in (\ref{redefinition}), 
and while it is natural to expect that the string field theory basis should 
exhibit `stringy' features most directly, it remains to be seen which form is 
more practicable for applications. 
We will leave these questions 
for future research.

\section*{Acknowledgments}
Helpful discussions and correspondence with Henriette Elvang, Dan Freedman, Chris Hull,
Seung Ki Kwak, Ashoke Sen, and especially Yu-tin Huang, Michael Kiermaier and Barton Zwiebach 
are gratefully acknowledged. I would also like to thank Princeton University for hospitality 
where this work has been finalized. 

This work is supported by the U.S. Department of Energy (DoE) under the cooperative
research agreement DE-FG02-05ER41360 and by the DFG -- The German Science Foundation.

\end{document}